\documentclass[english,12pt,aps,prd,a4paper,preprintnumbers,floatfix,nofootinbib,showpacs,superscriptaddress, notitlepage]{revtex4-1} 
 \pdfoutput=1
\usepackage[usenames,dvipsnames]{color}  
\usepackage{graphicx}
\usepackage{multirow}
\usepackage{rotating}
\usepackage{setspace}
\usepackage{braket}
\usepackage{caption}
\usepackage{subcaption}
\usepackage{amsmath}
\usepackage{amssymb}
\usepackage[colorlinks=true,citecolor=darkred,urlcolor=darkred, pdfborder={0 0 0}]{hyperref}
\usepackage[normalem]{ulem}
\usepackage{xcolor}
\usepackage{dsfont}
\usepackage{booktabs}
\usepackage{multirow}
\usepackage{tikz}		
\usepackage{adjustbox}		
\usepackage{enumitem}
\usepackage{mathtools}
\usepackage{booktabs}
\usepackage{feynmp-auto}
\usepackage{color}
 \usepackage{cancel}
\usepackage{hyperref}
\usepackage{lipsum}
\usepackage[T1]{fontenc}
\usepackage[utf8]{inputenc}
\usepackage[english]{babel}

\makeatletter
\def\p@subsection{}
\makeatother

\definecolor{darkred}{rgb}{0.6,0,0}

\definecolor{linkcolor}{rgb}{0,0,0.5}

\def\gsim{\raise0.3ex\hbox{$\;>$\kern-0.75em\raise-1.1ex\hbox{$\sim\;$}}}
\def\lsim{\raise0.3ex\hbox{$\;<$\kern-0.75em\raise-1.1ex\hbox{$\sim\;$}}}

\def\beqn#1{\begin{equation}\label{#1}}
\def\eeqn{\end{equation}}

\def\beqa#1{\begin{eqnarray}\label{#1}}
\def\eeqa{\end{eqnarray}}

\def\T{$\mathcal{T}$}

\def\Z2{$\mathcal{Z_2}$}

\def\vev#1{\left\langle #1\right\rangle}

\newcommand {\ignore}[1]{}

\newcommand{\sm}{{SM }}

\def\SM{$\mathrm{SU(3)_c \otimes SU(2)_L \otimes U(1)_Y}$}
\def\TrTrOne{ $\mathrm{SU(3)_c \otimes SU(3)_L \otimes U(1)_X } $}

\def\321{$\mathrm{SU(3) \otimes SU(2) \otimes U(1)}$}
\def\21{$\mathrm{ SU(2) \otimes U(1)}$}
\def\31{$\mathrm{SU(3) \otimes U(1)}$}

\newcommand{\AddrAHEP}{
  AHEP Group, Institut de F\'{i}sica Corpuscular --
  CSIC/Universitat de Val\`{e}ncia, Parc Cient\'ific de Paterna.\\
 C/ Catedr\'atico Jos\'e Beltr\'an, 2 E-46980 Paterna (Valencia) - SPAIN}

\newcommand{\AddrUnina}{Departimento di Fisica E. Pancini, 
Universit\`a di Napoli Federico II \\
Complesso Universitario di Monte Sant'Angelo, Via Cintia, Napoli (NA), Italy \\ and INFN, Sezione di Napoli.}

\newcommand{\AddINFN}{INFN, Sezione di Napoli, Complesso Universitario di Monte Sant'Angelo, Via Cintia, 80126 Napoli (NA), Italy}
 
\newcommand{\AddrIISERB}{Department of Physics, Indian Institute of Science Education and Research - Bhopal, \\ 
Bhopal Bypass Road, Bhauri, Bhopal 462066, India}

\begin{document}

\title{\color{BrickRed} Interpreting B-anomalies within an extended 331 gauge theory}
\author{Andrea Addazi}\email{Addazi@scu.edu.cn}
\affiliation{{\it Center for Theoretical Physics, College of Physics, Sichuan University, 610065 Chengdu, China} }
\affiliation{Laboratori Nazionali di Frascati INFN Via Enrico Fermi 54, I-00044 Frascati (Roma), Italy}
\author{Giulia Ricciardi}\email{giulia.ricciardi2@unina.it}
\affiliation{\AddrUnina}
\affiliation{\AddINFN}
\author{Simone Scarlatella}\email{s.scarlatella@studenti.unina.it}
\affiliation{\AddrUnina}
\author{Rahul Srivastava}\email{rahul@iiserb.ac.in}
\affiliation{\AddrIISERB}
\author{Jos\'{e} W. F. Valle}\email{valle@ific.uv.es}
\affiliation{\AddrAHEP}
 \begin{abstract}
   \vspace{1cm} 
   
In light of the recent $R_{K^{(*)}}$ data on neutral current flavour anomalies in $B \to  K^{(*)}  \ell^+ \ell^-$ decays,   we re-examine their quantitative interpretation in terms of an extended 331 gauge theory framework. We achieve this by adding two extra lepton species with novel 331 charges, while ensuring that the model remains anomaly free. In contrast to the canonical 331 models, the gauge charges of the first and second lepton families differ from each other, allowing lepton flavour universality violation. We further expand the model by adding the neutral fermions required to provide an adequate description for small neutrino masses.
\end{abstract}
\maketitle
\section{Introduction}
\label{sec:introduction}

In the past decade several measurements  of  $b$-quark decays with final leptons have shown disagreement with the overly successful Standard Model (SM).
Such disagreements are collectively referred to as “flavour anomalies”, and they typically feature tensions at the level of 2–3 standard deviations between experimental results and SM predictions.
An interesting aspect of these anomalies lies in the fact that they all seem to point towards the presence of lepton flavour universality (LFU) violation in the interactions mediating the processes. 
Last year, the measurements of rare decays $B^+ \to  K^+ \ell^+ \ell^-$, with $\ell$ denoting an electron or a muon,
have provided the further evidence for the breaking of LFU in beauty-quark decays
in a single process, with a significance of 3.1 standard deviations, based on 9 fb$^{-1}$ of proton-proton collision data collected at LHCb \cite{LHCb:2021trn}.

The accuracy of the  predictions for the branching fractions of  semileptonic $B$ decays is generally higher than the one of hadronic decays, due to the reliability of perturbative techniques. 
Moreover, this precision can be further increased by taking ratios of processes with electrons or muons in the final state, since they are affected equally by the strong force, which does not couple
directly to leptons. 
Thus to minimize the hadronic uncertainties one usually introduces branching fraction ratios  which in the case of  $B \to  K^{(*)} \ell^+ \ell^-$ can be defined as 
\begin{equation}\begin{split}
    &R_{K^{(*)}[q^2_{\rm min},q^2_{\rm max}]}=\frac{\mathcal B(B\to K^{(*)} \mu^+\mu^-)_{q^2\in[q^2_{\rm min},q^2_{\rm max}]}}{\mathcal B(B\to K^{(*) }e^+e^-)_{q^2\in[q^2_{\rm min},q^2_{\rm max}]}}
  \end{split}\end{equation}
where $\mathcal{B}$ denotes the branching fraction for the given decay mode measured over a bin size of $[q^2_{\rm min},q^2_{\rm max}]$. 
The resulting $R_{K^{(*)}}$ are measured over specific ranges for the squared di-lepton invariant mass $q^2$. 

The $B \to  K^{(*)} \ell^+ \ell^-$ decays are driven  at the  quark-level by the $b \to  s\ell^+ \ell^-$ decay.
The hadronic process involved is mediated by Flavour Changing Neutral Currents (FCNCs), which are forbidden at tree-level in the SM.
The branching fractions in the ratio $R_{K^{(*)}}$ differ only by the leptons in the final state, hence this ratio is expected to be 1 by virtue of Lepton Flavour Universality (LFU),
with small deviations induced by phase space differences and QED corrections. 
By comparing   recent LHCb experimental values with  theoretical determinations we have: 
\begin{widetext}
\begin{equation}
\begin{aligned}[l]
	&R_{K^+[1.1,6.0]}^\text{exp}=0.846^{+0.042\, +0.013}_{-0.039\,-0.012}\text{~\cite{LHCb:2021trn}}\\
	&R_{K^{*0}[0.045,1.1]}^\text{exp}=0.66^{+0.11}_{-0.07}\pm 0.03\text{~\cite{LHCb:2017avl}}\\
	&R_{K^{*0}[1.1,6.0]}^\text{exp}=0.69^{+0.11}_{-0.07}\pm 0.05\text{~\cite{LHCb:2017avl}}\\
\end{aligned}
\quad
\begin{aligned}[l]
	&R_{K^+}^\text{th}=1.00\pm 0.01\text{~\cite{Bordone:2016gaq,Capdevila:2017ert}}\\
	&R_{K^{*0}[0.045,1.1]}^\text{th}=0.922\pm 0.022\text{~\cite{Capdevila:2017ert}}\\
	&R_{K^{*0}[1.1,6.0]}^\text{th}=1.000\pm 0.006\text{~\cite{Capdevila:2017ert}}\\
\end{aligned}
\quad
\begin{aligned}[l]
        & 3.1~ \sigma\\
        & 2.3~ \sigma\\
        & 3.4~ \sigma\\
\end{aligned}
\label{lhcbdata}
\end{equation}\end{widetext}
where $q^2$ is given in GeV$^2$. In the experimental data the first errors are statistical and the second ones systematic. %
The first result is the most precise measurement to date and consistent with the SM prediction with a p-value of 0.10\%.
This gives evidence for the violation of lepton universality in these decays with a significance of 3.1$\sigma$.
We have also listed the statistical significance of the anomalies for the other experimental results.

Recently LHCb has investigated $B^0 \to  K^{0}_S\ell^+ \ell^-$ and $B^+ \to  K^{*+} \ell^+ \ell^-$ decays, with $\ell$ being an electron or a muon. 
Notice that these decays involve mesons which are the isospin partners of the ones in the previously measured channels $B^+ \to  K^{+} \ell^+ \ell^-$  and  $B^0 \to  K^{*0} \ell^+ \ell^-$.
Although these decays have similar branching fractions as their isospin partners, they suffer from a reduced experimental efficiency at LHCb, due to the presence of a long-lived $K^{0}_S$ or $\pi^0$
meson in the final states. The measured ratios are 
\begin{widetext}
\begin{equation}
\begin{aligned}[l]
&R_{{K^{0}_S}[1.1,6.0]}^\text{exp}=0.66^{+0.20\, +0.02}_{-0.14\,-0.04}\text{~\cite{LHCb:2021lvy}}\\
&R_{K^{*+}[0.045,6.0]}^\text{exp}=0.70^{+0.18\, +0.03}_{-0.13\,-0.04}\text{~\cite{LHCb:2021lvy}}\\
\end{aligned}
\label{lhcbdataoct}
\end{equation}\end{widetext}
and provide $\sim$ 1.5$\sigma$ hints of departures from the SM~\cite{LHCb:2021lvy}. 
  
Recent experimental determinations of  $R_{K^{*}}$ have also been given by the Belle collaboration, using the full $\Upsilon(4S)$ data sample containing 772 $\times 10^6$ $B \bar B$ events.
For the same range of di-lepton invariant mass reported in \eqref{lhcbdata}, they find 
\begin{widetext}
\begin{equation}
\begin{aligned}[l]
	&R_{K^{*0}[0.045,1.1]}^\text{exp}=0.46^{+0.55}_{-0.27}\pm 0.13\text{~\cite{Belle:2019oag}}\\
	&R_{K^{*0}[1.1,6.0]}^\text{exp}=1.06^{+0.63}_{-0.38}\pm 0.14\text{~\cite{Belle:2019oag}}\\
	&R_{K^{*+}[0.045,6.0]}^\text{exp}=0.62^{+0.60}_{-0.36}\pm 0.09\text{~\cite{Belle:2019oag}}\\
\end{aligned}
\quad
\end{equation}\end{widetext}
Babar, Belle and LHCb have provided other prominent contributions to ratio determinations in these as well as in different
channels \cite{BaBar:2012obs,BaBar:2013mob,LHCb:2015gmp,Belle:2015qfa, Belle:2016kgw, LHCb:2014vgu, LHCb:2017avl, Belle:2019oag, LHCb:2021lvy}.

The primary requirement for any model to explain these $b \to s$ anomalies is to have a symmetry which distinguishes between semi-leptonic $B$ decays
to $\mu^+ \mu^-$ and to $e^+ e^-$ such that $R_{K^{(*)}}$ deviates appreciably from one.
 Within the \sm this cannot be achieved as the theory is sequential, so that the $e^-$ and $\mu^-$ carry the same gauge charges.
 One possible way out is  to postulate a new $U(1)_X$ gauge symmetry under which $e^-$ and $\mu^-$ carry different
 charges~\cite{Altmannshofer:2016jzy,Crivellin:2015era,Bonilla:2017lsq,Allanach:2020kss,Bause:2021prv}.

 Here we propose that the deviation from $R_{K^{(*)}} = 1$ can be achieved from a bigger non-trivial gauge symmetry,
 from which the standard \SM~gauge group emerges as a subgroup.  
We do this in the framework of the so-called 331 models~\cite{Singer:1980sw,Valle:1983dk,Montero:1992jk}, which constitute one of the simplest well-motivated extensions of the SM.
The name 331 follows from their extended \TrTrOne~gauge group.
Several issues which remain unanswered within the SM, for instance the origin of light-neutrino masses, typically call for larger gauge structures and/or new particles.
Grand Unified Theories play certainly an important role in this respect, but 331 models have the advantage that they can provide scenarios
where larger gauge symmetries can be probed already at the TeV scale.
Moreover, so far no hard evidence in favour of conventional unification schemes has been found. 
Here we note that 331 models lead to a consistent theoretical structure and also a phenomenologically viable weak neutral
current~\footnote{Earlier 331 models were suggested to account for the high y-anomaly,
  which turned out to be a fake~\cite{Lee:1977qs,Lee:1977tx,Buccella:1977gx,Buccella:1978nc}.
  while shedding light on mysteries such as the number of particle families. }\textsuperscript{,}\footnote{Recently there are some attempts to relate the $(g-2)_{\mu}$ anomaly with 331 models
    (see for Refs.\cite{deJesus:2020ngn,Hue:2020wnn,Hue:2021zyw}).}.
  As a result, 331-based extensions have attracted a lot of interest, see for
  instance~\cite{Boucenna:2014dia, Dong:2014wsa, Alves:2016fqe, Dias:2020kbj, Hernandez:2021zje}. 
These models experience two stages of breaking: at a larger scale $\Lambda_{NP}$, the extended group is broken down to the SM gauge group,
while the electroweak symmetry breaking occurs at the lower scale $\Lambda_{EW}$.
Phenomenologically, these models feature additional heavy gauge bosons,
as well as an extended Higgs sector to drive the two spontaneous symmetry breakdowns.

Left-handed fermions transform according to one of the two fundamental representations, i.e. triplets (or antitriplets) under the action of $SU(3)_L$. 
In the simplest version of 331 theories~\cite{Singer:1980sw,Valle:1983dk} exactly three families emerge from the cancellation of chiral anomalies,
  which requires that the number of triplets matches the number of antitriplets.
In contrast to the SM, where the anomaly is canceled within each generation of fermions, in these 331 models all families must be considered to fulfill the anomaly cancelation. 
Since quarks come in three colours, there must be three families of quarks and leptons, with leptons appearing in the same fundamental representation of the group.
As a result their couplings with gauge bosons are necessarily family-independent, preventing any LFU violation in their gauge couplings. 

Here we are concerned with other versions of the 331 model extending the lepton sector with additional species. 
This assumption allows us to choose at least one lepton  family transforming differently from the others, ensuring the presence of LFU violation. 
The minimal choice preserving anomaly cancellation requires two additional lepton species.
These versions of the 331 model have been considered in Refs.~\cite{Cabarcas:2012uf, Cabarcas:2013jba,Diaz:2004fs,Diaz:2003dk, Ponce:2001jn,Anderson:2005ab}.
In the  preliminary analysis  \cite{Descotes-Genon:2017ptp}, it was studied whether they can reproduce the anomalies observed in $b\to s\ell\ell$ processes under simple assumptions: 
LFU violation is dominated by neutral gauge boson exchange, with no significant Lepton Flavour Violation (LFV) of the form
$b\to s\ell_1\ell_2$, nor large contributions to $B_s\bar{B}_s$ mixing. 
It was found  that under these simple assumptions an extended 331 model without exotic electric charges for fermions and gauge bosons can yield large contributions to $(C_9^\mu,C_{10}^\mu)$
in good agreement with 2018 global fit analyses \cite{Capdevila:2017bsm}.
This result is rather non-trivial, given that the model is quite constrained.

Apart from providing an updated numerical analysis including the recent $B$-anomaly data, here we fully develop the proposal in Ref. \cite{Descotes-Genon:2017ptp},
by adding the neutral fermions required for an adequate description of the neutrino mass matrix.
In addition to gauge symmetries, we assume the presence of two auxiliary discreet  $\mathbb{Z}_2$ and $\mathbb{Z}_3$ symmetries, which are needed in order to ensure an adequate pattern of fermion masses. As we describe in Sec. \ref{sec:model} in more details, the primary purpose of the $\mathbb{Z}_3$ symmetry  is to forbid direct gauge invariant couplings between SM and exotic leptons. Presence of such coupling would imply either unacceptable large masses for SM leptons or unacceptable small masses for exotic charged leptons, both scenarios of course being experimentally rejected. 
An additional $\mathbb{Z}_2$ is further needed to generate different masses for  SM and exotic fermions which carry the same gauge quantum numbers, without need for fine tunning.

The paper is organized as follows. In Sec.~\ref{sec:model} we sketch the model and its field representations.
In Sec.~\ref{sec:yukawa-interactions} we discuss the Yukawa interactions, including those used in the implementation of the seesaw mechanism.
In Sec.~\ref{sec:ferm-mass-matr} we comment on fermion mass generation, including neutrino masses.
In Sec.~\ref{sec:b-flavour-global} we perform a comparison with $B$ flavour global analyses.
We found that this 331 model can generate large new physics contributions to $(C_9^\mu,C_{10}^\mu)$  parameters, in agreement with new physics  scenarios 
favoured by global fits. In Sec.~\ref{Conclusion} we present our conclusions.

\vspace{-0.2cm}
\section{The Model}
\label{sec:model}

Apart from gluons, any 331 model has  nine vector bosons associated to each generator of the gauge group, eight $W^a_\mu$ for SU(3)$_\text{L}$ and one $X_\mu$ for U(1)$_\text{X}$. 
We indicate the generators of the SU(3)$_\text{L}$ gauge group  with $\hat T^1 \cdots \hat T^8$, normalized as $\mathrm{Tr}[\hat T^i \, \hat T^j]=\delta^{ij}/2 $,
and define the $U(1)_X$ generator as $\hat T^9 = {\mathds 1}/\sqrt{6}$, where  ${\mathds 1} = \mathrm{diag} (1, 1, 1)$  is the identity matrix. 
The electric charge is defined in general as a linear combination of the diagonal generators of the group
\begin{equation}
\hat Q  = a \hat T^3+ \beta \hat T^8+X{\mathds 1} \end{equation}
where the values of the proportionality constants $a$ and $\beta$ distinguish different 331 models. 
We have  $\hat T^3 = 1/2 \, \hat{\lambda}^3= 1/2 \, \mathrm{diag}(1, -1,0)$ and $\hat T^8 = 1/2 \, \hat{\lambda}^8= 1/(2\sqrt{3}) \, \mathrm{diag}(1, 1,-2)$,
where $\hat{\lambda}^i$ are the Gell-Mann matrices. $X$ is the quantum number associated with $U(1)_X$. 
We set $a=1$ to obtain isospin doublets which embed \21 into \31.
In order to restrict $\beta$ we demand that no new particle introduced in the model has exotic charges (i.e. different from the SM ones).
This can be done by choosing the particular value
\begin{equation}
  \beta= -1/\sqrt3
\end{equation}
which is the original assignment made in~\cite{Singer:1980sw}. We will thus have the following definition of the electric charge operator
 \begin{equation}
   \hat Q=  \hat T^3-\frac{1}{\sqrt3} \hat T^8+X{\mathds 1}
\label{charge1}
\end{equation}
Complex gauge fields are defined by the combinations 
$W^{\pm}_{\mu}=\frac{1}{\sqrt{2}}(W^1_{\mu}{\mp}iW^2_\mu)$,
$V^{\pm}_{\mu}=\frac{1}{\sqrt{2}}(W^6_{\mu}{\mp}iW^7_\mu)$ and 
$Y^{0(0_*)}_{\mu}=\frac{1}{\sqrt{2}}(W^4_{\mu}{\mp}iW^5_\mu)$,
where the superscripts $\pm,0$ denote electric charges of the fields, a notation we will follow throughout this work.
In general, the values of the electric charges of the $V_\mu$ and $Y_\mu$ bosons depend on the value of $\beta$. 
With our choice of $\beta=-\frac{1}{\sqrt{3}}$ the electric charges of all gauge bosons are fixed to either $\pm1$ or $0$, i.e. non-anomalous values. 

 \subsection{Symmetry breaking}

Starting from the $\mathrm{SU(3)_c \otimes SU(3)_L \otimes U(1)_X }  $ gauge group (with gauge couplings $g_S,g,g_X$), the model will undergo two spontaneous symmetry breakings (SSB)
triggered by colour singlet scalar fields acquiring non-vanishing vacuum expectation values, in a way analogous to the SM.
The overall pattern of SSB is the following 
\vspace{3mm}
\begin{widetext}
\begin{center}
\begin{adjustbox}{max width=\textwidth}
\begin{tikzpicture}
\node at (0,0) {$SU(3)_\text{c}\times SU(3)_\text{L} \times U(1)_X$};
\draw [->] (2.5,0) --  node[above] {}node[below] {$\Lambda_\text{NP}$}  (3.5,0);
\node at (6,0) {$SU(3)_\text{c}\times SU(2)_\text{L} \times U(1)_Y$};
\draw [->] (8.5,0) --  node[above] {}node[below] {$\Lambda_\text{EW}$}  (9.5,0);
\node at (11.5,0) {$SU(3)_\text{c}\times U(1)_\text{EM}$}; 
\end{tikzpicture}
\end{adjustbox}
\end{center}\end{widetext}
The first SSB occurs at an energy scale $\Lambda_\text{NP}$ and allows to recover the \sm gauge group.
The subsequent one, at energy scale $\Lambda_\text{EW}$, reproduces the electroweak symmetry breaking (EWSB) of the SM. 
We assume that $\Lambda_\text{NP}\gg\Lambda_\text{EW}$, and introduce a small parameter $\epsilon=\Lambda_\text{EW}/\Lambda_\text{NP}$ characterizing the order of magnitude of the new physics (NP).

As in the SM, the Higgs fields, besides giving mass to the gauge bosons, are used to generate fermion mass terms through gauge invariant Yukawa terms. 
The need to build gauge invariant terms in such a way to obtain appropriate mass terms after SSB constrains possible scalar Higgs field representations. 
Since the fermions transform either as a $3$ or as a $\bar 3$ under SU(3)$_{\textrm{L}}$, we only have a limited number of possibilities~\cite{Diaz:2003dk} for a scalar field $\Phi$,
which at both stages can only be a triplet, a sextet or a singlet~\footnote{A 331 gauge singlet scalar can in principle contribute to the neutral fermion mass term;
  however, since it does not  change our conclusions, we ignore this possibility for simplicity. Though we have a different number of  triplets and  sextets than in Ref.~\cite{Diaz:2003dk},
  their conclusions on the structure of the gauge boson mass sector do not change, since we assume the same vacuum expectation value (vev) alignments.}.
    
We assume that the breaking of the SU(3)$_\text{L}$ symmetry is accomplished through two triplets $\chi$ and $\tilde \chi$ and a sextet  $S_1$.
There are five gauge fields that acquire a mass of the order of $\Lambda_\text{NP}$, whereas the remaining three gauge fields are the SM gauge bosons.  
At the first SSB stage, the gauge bosons acquiring mass are the charged ones $V^{\pm}$, the neutral gauge bosons,  $Y^{0(0_*)}$ and a massive neutral gauge boson $Z'$ given as a combination of the
two neutral gauge bosons $X, W^8$, which also yields the gauge boson $B$. Their mixing angle $\theta_{331}$ is given by:
\begin{equation}
\begin{pmatrix}Z'\\B\end{pmatrix}=\begin{pmatrix}\cos\theta_{331}&-\sin\theta_{331}\\\sin\theta_{331}&\cos\theta_{331}\end{pmatrix}\begin{pmatrix}X\\W^8\end{pmatrix},
\end{equation}
The angle $\theta_{331}$ is found by singling out the $Z'$ field in the sector of the Lagrangian including the masses of the gauge bosons,
which follow from the covariant derivative in the Higgs Lagrangian. It yields
\begin{equation}
\sin\theta_{331}=\frac{g}{\sqrt{g^2+\frac{g_X^2}{18}}}\,,\qquad
\cos\theta_{331}=-\frac{\frac{g_X}{3\sqrt2}}{\sqrt{g^2+\frac{g_X^2}{18}}}.
\label{mixing:angle}
\end{equation}
where $g$,$g_X$ denote the coupling constants for $SU(3)_L$ and $U(1)_X$ respectively.

The second stage of symmetry breaking is the usual electroweak symmetry breaking to the electromagnetic subgroup i.e. \SM $\to \text{SU(3)}_\text{c} \otimes \text{U(1)}_\text{EM}$.  
This breaking is driven by the 
triplets $\eta, \rho$, $\tilde \eta, \tilde \rho$ 
and the sextet $S_{c}$. 
After electroweak symmetry breaking, the neutral gauge bosons $W^3$ and $B$ mix with each other to give the \sm $Z$ and $\gamma$ bosons as follows
\begin{equation}
\begin{pmatrix} Z\\ \gamma \end{pmatrix}=\begin{pmatrix}\cos\theta_{W}&-\sin\theta_{W}\\\sin\theta_{W}&\cos\theta_{W}\end{pmatrix}\begin{pmatrix} W^3 \\ B \end{pmatrix},
\end{equation}
where the mixing angle $\theta_{W}$ is the usual electroweak mixing angle. \\

Summarizing, our scalar sector is similar to that in Ref. \cite{Descotes-Genon:2017ptp}, except for the addition of the triplets $\tilde \chi$, $\tilde \eta, \tilde \rho$
  and the removal of the sextet $S_b$, for reasons that will be detailed later. 
  The two-step spontaneous symmetry breaking ensures that all the new gauge bosons indeed get large masses through the large vevs of the
  scalars breaking \TrTrOne $\to$ \SM.
  Only the SM gauge bosons get their masses in the second symmetry breaking step 
  due to the electroweak-scale vev carried by the scalars breaking \SM $\to \text{SU(3)}_\text{c} \otimes \text{U(1)}_\text{EM}$. 

 \subsection{ Matter content} 
\label{sec:fieldsandrepr}

In the previous  section we have discussed the gauge structure and symmetry breaking pattern, here we focus on the matter content of our 331 model, looking into detail of the charge assignment. 
This 331 model  contains three families of left-handed quarks and five families of left-handed leptons~\cite{Ponce:2001jn,Anderson:2005ab,Cabarcas:2012uf,Cabarcas:2013jba,Descotes-Genon:2017ptp}.
They all belong to the fundamental representations of SU(3)$_\text{L}$. Two generations of quarks and one of leptons behave as anti-triplets, all the others as triplets of SU(3)$_\text{L}$. 
This  fermion content  ensures at the same time the cancellation of the anomalies and allows LFU violation, but otherwise departs from the SM as little as possible. 
Fixing $\beta=-1/\sqrt{3}$ has ensured  that both SM and new fields in the spectra all have non-exotic charges. 

Using the notation (SU(3)$_\text c$, SU(3)$_\text{L}$, U$_X$(1)) while referring to the representations of the fermions, we write for the left-handed ones 
\begin{itemize}
\item three families of quarks~\footnote{Note that the order in which the triplet components are arranged is a matter of choice. An alternative convention is to have the first component of
    quark triplets to be up-type, whereas the others are down-type. For leptons the upper one would be charged, while the others neutral. The third component is always exotic~\cite{Singer:1980sw}. }
\begin {equation}
\begin{split}
 q_m &=\begin{pmatrix}d^L_m\\-u^L_m\\B^L_m\end{pmatrix}\sim (3, \bar 3, 0), \quad m=1,2 \\
q_3 &=\begin{pmatrix}u^L_3\\d^L_3\\T^L_3\end{pmatrix}\sim (3, 3, \frac 1 3);
\end{split}
\label{qh}
\end{equation}
\item five species of leptons
\begin {equation}
\begin{split}
\ell_1&=\begin{pmatrix}e^{-L}_1 \\ -\nu^L_1 \\ 
E^{-L}_1\end{pmatrix}\sim (1, \bar3, -\frac 2 3), \\
\ell_n &=\begin{pmatrix}\nu^L_n\\ e^{-L}_n \\N^{0L}_n\end{pmatrix}\sim (1, 3, -\frac 1 3), \qquad n=2,3 \\
L_4 &=\begin{pmatrix} \nu^{0L}_4\\ E^{-L}_4 \\ N^{0L}_4\end{pmatrix}\sim (1, 3, -\frac 1 3),  \\ 
L_5 &=\begin{pmatrix}\bigl(E^{-R}_4\bigr)^c\\ N^{0L}_5 \\ \bigl(e^{-R}_3\bigr)^c\end{pmatrix}\sim (1, 3, \frac 2 3). \\
\end{split}
\label{lh}
\end{equation}
\end{itemize} 

Notice that, as in the original 331-model in~\cite{Singer:1980sw}, no positively charged leptons have been introduced in the triplets.
 Indeed, they would only appear in  $L_5$, but we identify them with the charge conjugate of the right-handed components of $E^{-}_4$ and  $e^{-}_3$.
  This  economical identification avoids the presence of charged exotic particles at the electroweak scale.
We have labelled the SM fermions with lower-case ($e_i, \, \nu_i$ with $i=1,2,3$), and the exotic ones with $\nu_4$ and upper-case ($E_{1,4}, \, N_{2,3,4,5}$), choosing letters and/or superscripts recalling their electric charge assignments and chirality. 
In contrast, reference to chirality has been eliminated for simplicity when naming left-handed triplets/antitriplets as a  whole:
left-handed SM quarks, SM leptons and exotic leptons are indicated with $q_{1,2,3}$, $\ell_{1,2,3}$ and $L_{4,5}$, respectively. Capital letters have been used for the last two triplets because they include only exotic fermions.

The right-handed components of charged fermions are defined as singlets of SU(3)$_\text{L}$; the SM ones are labelled as $u_{1,2,3}$, $d_{1,2,3}$ and $e_{1,2}$ with lower-case,
and the exotic ones $B_{1,2}$, $T_{3}$ and $E_{1}$ with upper-case, without any chirality or charge superscript.
Altogether, we have the following list of right-handed fermions 
\begin{itemize}
\item the quark fields
\begin {equation}
\begin{split}
d_{1,2,3}  &\sim(3,1,-1/3)\\  
B_m  &\sim(3,1,-1/3),\qquad m=1,2\\
u_{1,2,3} &\sim(3,1,2/3)\\  
T_3 &\sim(3,1,2/3)
\end{split}
\end{equation}
\item the charged lepton fields
\begin {equation}
\begin{split}
e_{1,2} &\sim(1,1,-1) \\
 E_{1} &\sim(1,1,-1)
\end{split}
\label{rh}
\end{equation}
As already mentioned, the right-handed parts of $e_3^-$ and $E_4^-$ are included in the SU(3)$_\text L$ lepton triplet $L_5$. 
\item the neutral lepton fields \footnote{Compared with the fermion content of Ref. \cite{Descotes-Genon:2017ptp}, we have three extra neutral two-component fermions $\nu^R_{1,2,3}$
      to implement neutrino mass generation \textit{\`a la seesaw}.}
\begin {equation}
\begin{split}
\nu^R_{1,2,3} \sim (1,1,0) 
\end{split}
\end{equation}
We do not include right-handed partners for the neutral lepton fields $N^{0L}_{2,3,4,5}$ and $\nu^{0L}_4$, which get Majorana mass terms. 

\end{itemize}

The representation assigments for the fermions and scalars are summarized in Table~\ref{tab:seesaw-z2}, where one also sees the presence of two auxiliary discrete
symmetries $\mathbb{Z}_2$ and $\mathbb{Z}_3$. 
The latter is the discrete abelian cyclic group of order 3. It has three elements and a convenient representation is obtained by using the cube roots of unity.
These are given by $1, \omega, \omega^2$ where $\omega = exp[\frac{2\pi i}{3}]$ with $\omega^3 =1$.
Note that $\omega^{-1} = \omega^2$ and that $\omega^{3n} = 1$ if  $n$ is an integer.
This cyclic nature further implies that $\omega^n = \omega^{n-3}$, so that $\omega^4 = \omega^3 \times \omega = \omega$, $\omega^5 = \omega^3 \times \omega^2 = \omega^2$ and so on.
 These extra symmetries are needed in order to ensure an adequate pattern of fermion masses.
 In the absence of the $\mathbb{Z}_3$ symmetry, e.g., the unwanted invariant mass term $\bar{\ell}_1 (L_5)^c$ would be present.
  On the other hand, since the SU(3)$_\text c$, SU(3)$_\text L$, U$_X$(1) gauge charge as well as the  $\mathbb{Z}_3$ charges of the SM fermion triplets $\ell_{2,3}$ and of
    the exotic triplet $L_{4}$ are the same, these symmetries cannot distinguish between the SM and the exotic fermions inside the $L_4$ triplet.
    To prevent having similar masses for the exotic and SM fermions, we make a distinction between them by means of an additional $\mathbb{Z}_2$ symmetry, as shown in Table. \ref{tab:seesaw-z2}. 

\begin{table}[!t]
\centering
\begin{tabular}{| c || c | c | c | c || c | c | c  | c | }
  \hline 
& Fields    &    $\rm SU(3)_c \otimes SU(3)_L \otimes U(1)_X$   &\hspace{.05cm}  $\mathbb{Z}_3$ \hspace{.05cm}   &\hspace{.05cm}  $\mathbb{Z}_2$ \hspace{.05cm}    
& Fields    &    $\rm SU(3)_c \otimes SU(3)_L \otimes U(1)_X$   &\hspace{.05cm}  $\mathbb{Z}_3$ \hspace{.05cm}   &\hspace{.05cm}  $\mathbb{Z}_2$ \hspace{.05cm}     \\
\hline \hline
\multirow{4}{*}{ \begin{turn}{90} \hspace{0.9cm} \small{Quarks} \hspace{0.05cm} \end{turn} }
& $q_{1,2}$   & ($\mathbf{3}, \mathbf{\bar{3}}, \mathbf{0}$)  & $\mathbf{1}$  &  $\mathbf{1} $ 
& $q_3$   & ($\mathbf{3}, \mathbf{3}, \mathbf{1/3}$)      & $ \mathbf{1}$ &  $\mathbf{1} $      \\
& $u_{1,2,3}$   & ($\mathbf{3}, \mathbf{1}, \mathbf{2/3}$)    & $ \mathbf{\omega^2}$  &  $\mathbf{1} $      
& $d_{1,2,3}$   & ($\mathbf{3}, \mathbf{1}, \mathbf{-1/3}$   & $ \mathbf{\omega} $   &  $\mathbf{1} $\\
& $T_3$   & ($\mathbf{3}, \mathbf{1}, \mathbf{2/3}$)  & $\mathbf{\omega^2}$ &  $\mathbf{1}$    
& $B_{1,2}$   & ($\mathbf{3}, \mathbf{1}, \mathbf{-1/3}$) & $\mathbf{\omega}$  & $\mathbf{1}$    \\
\hline \hline
\multirow{4}{*}{ \begin{turn}{90} \hspace{0.9cm} \small{Leptons}  \hspace{0.25cm} \end{turn} }
& $\ell_1$ & ($\mathbf{1}, \mathbf{\bar{3}}, \mathbf{-2/3}$)  & $\mathbf{1}$  & $\mathbf{1}$ 
& $\ell_{2,3}$   & ($\mathbf{1}, \mathbf{3}, \mathbf{-1/3}$)  &  $ \mathbf{\omega}$  &  $\mathbf{1} $     \\
& $e_{1,2}$   & ($\mathbf{1}, \mathbf{1}, \mathbf{-1}$)   &  $\mathbf{\omega}$ &  $\mathbf{1} $ 
& $E_1$   & ($\mathbf{1}, \mathbf{1}, \mathbf{-1}$)   &  $\mathbf{\omega}$ &  $\mathbf{-1}$        \\
& $L_4$  & ($\mathbf{1}, \mathbf{3}, \mathbf{-1/3}$)  &  $\mathbf{\omega}$  & $\mathbf{-1}$  
& $L_5$  & ($\mathbf{1}, \mathbf{3}, \mathbf{2/3}$)   &  $\mathbf{\omega}$  & $\mathbf{-1}$  \\
& $\nu^R_{1,2,3}$ 	  &  ($\mathbf{1}, \mathbf{1}, \mathbf{0}$)    &  $\mathbf{1}$    &  $\mathbf{1} $ 
&                     &                                            &                  &    \\
\hline \hline
\multirow{5}{*}{ \begin{turn}{90}\hspace{1.5cm} \small{Scalars} \hspace{0.25cm} \end{turn} }
& $\chi$   &  ($\mathbf{1}, \mathbf{3}, \mathbf{-1/3}$) &  $\mathbf{\omega}$   &  $\mathbf{1} $         
& $S_1$  &  ($\mathbf{1}, \mathbf{6}, \mathbf{-2/3}$) &  $ \mathbf{\omega^2}$ & $\mathbf{1} $    \\  
& $\tilde{\chi}$   &  ($\mathbf{1}, \mathbf{3}, \mathbf{-1/3}$) &  $\mathbf{\omega}$   &  $\mathbf{-1} $        
& $\tilde{\eta}$   &  ($\mathbf{1}, \mathbf{3}, \mathbf{-1/3}$)   &  $\mathbf{\omega}$   & $\mathbf{-1}$  \\
&$\eta$   &  ($\mathbf{1}, \mathbf{3}, \mathbf{-1/3}$)   &  $\mathbf{\omega}$   &  $\mathbf{1}$       
&$\rho$  &  ($\mathbf{1}, \mathbf{3}, \mathbf{2/3}$)    &  $\mathbf{\omega^2}$ &  $\mathbf{1}$   \\
& $S_c$   &  ($\mathbf{1}, \mathbf{6}, \mathbf{4/3}$)    &  $\mathbf{\omega^2}$ &  $\mathbf{1}$          
& $\tilde{\rho}$      & ($\mathbf{1}, \mathbf{3}, \mathbf{2/3}$)  &  $ \mathbf{1} $  &  $\mathbf{1} $   \\
    \hline
  \end{tabular}
  \caption{\begin{footnotesize} Particle content 
  of the  331 model, 
    where  in addition to the SU(3)$_\text c$, SU(3)$_\text L$, U$_X$(1) gauge symmetries, we have listed two abelian discrete symmetries, see text. 
 \end{footnotesize}}
 \label{tab:seesaw-z2}
\end{table}

\vspace{-.55cm}
\section{Yukawa interactions}
\label{sec:yukawa-interactions}

Before discussing the details of the fermion masses, we summarize the Higgs scalar representations what will drive the breaking of $\rm SU(3)_c \otimes SU(3)_L \otimes U(1)_X$ in the Yukawa sector ~\cite{Diaz:2003dk, Descotes-Genon:2017ptp}.
There are two stages of symmetry breaking, at the high 331 scale and the EW scale. Vevs of a generic field $\psi$ are denoted by $\langle\psi  \rangle$.

\subsection{331 Breaking}
\label{sec:331-breaking}

This is the first  SSB stage,  which is accomplished by the $\rm SU(3)_L$ scalar sextet $S_1$ and triplets $\chi,  \tilde{\chi}$ with (U(1)$_\mathrm{X}, \mathbb{Z}_3, \mathbb{Z}_2)$ charges and non-zero vevs as follows: 
\begin{equation}\begin{split}
\langle S_1\rangle&=\begin{pmatrix} 	0&0&0\\
			0&0&0\\
			0&0& \langle (S_1)_{33} \rangle
\end{pmatrix},\,(\mathrm{U(1)}_\mathrm{X}, \mathbb{Z}_3, \mathbb{Z}_2) = (-\frac 2 3, \omega^2, 1)
\\
\langle \chi\rangle&=\frac 1 {\sqrt 2} \begin{pmatrix} 0\\0\\ \langle \chi_3 \rangle \end{pmatrix},\,(\mathrm{U(1)}_\mathrm{X}, \mathbb{Z}_3, \mathbb{Z}_2) = (-\frac 1 3, \omega, 1)
\\
\langle \tilde{\chi}\rangle&=\frac 1 {\sqrt 2} \begin{pmatrix} 0\\0\\ \langle\tilde{\chi} \rangle\end{pmatrix},\,(\mathrm{U(1)}_\mathrm{X}, \mathbb{Z}_3, \mathbb{Z}_2) = (-\frac 1 3, \omega, -1)
\end{split}
\end{equation}
The $\mathbb{Z}_3 \otimes \mathbb{Z}_2$ and gauge symmetry invariant Yukawa terms that can be built with the sextet are: 
\begin {eqnarray}
	&&\bar\ell_a S_1(\ell_b)^c\quad \qquad  a,b = 2,3 \nonumber \\ &&	\bar{L}_4 S_1(L_4)^c  
\end{eqnarray}
These terms lead  to Majorana masses for the exotic neutral leptons $ N^0_{2,3,4}$.

The $\mathbb{Z}_3 \otimes \mathbb{Z}_2$ and gauge symmetry invariant Yukawa terms 
that can be built with the triplets are:  
\begin{itemize}
\item 
The up- and down-quark mass terms  
  \begin{equation}
	\begin{split}
	&  \bar{q}_m \chi^* D \quad \qquad  m = 1,2  \\
		&\bar{q}_3 \chi\,  U
		\end{split}
	\end{equation}
        where $D$ represents any right-handed $d_{1,2,3}$ or $B_{1,2}$, while $U$ represents any right-handed $u_{1,2,3}$ or $T_3$. 
        After SSB, they contribute to mix charged SM and exotic quarks, and give Dirac mass to $B_{1,2}$ and $T_3$.
\item
The equivalent terms in the lepton sector  
 \begin{equation}
 \begin{split}
	& \bar{\ell}_1 \chi^*  e_{1} \\
		&  \bar{\ell}_1 \chi^* e_{2} \\
		&  \bar{\ell}_1 \tilde{\chi}^* E_1
		\end{split}
	\end{equation}
Here one sees how the scalar triplet $\tilde{\chi}$, odd under the $\mathbb{Z}_2$ symmetry, allows a coupling between $E_1$ with $\ell_1$, providing a Dirac mass term for $E_1$. 
\item

  We also have the anti-symmetric combination of SU(3)$_\text{L}$ triplets or antitriplets,  i.e. 
 \begin{equation} 
 \begin{split}
 & \epsilon_{ijk}\chi^{*i}\bar{L}^{j}_4(L_5)^{c\,k}  \\
&  \epsilon_{ijk}\tilde{\chi}^{*i}\bar\ell^{j}_{m}(L_5)^{c\,k} \quad \qquad  m = 2,3
\end{split}
	\end{equation}
where the  $i,j,k=1,2,3$ indices refer to SU(3)$_\text{L}$.
The first term includes mixing between  $N^0_5$ and $\nu^{0L}_4$ and allows mass term for $E_4$. 
\end{itemize}

Summarizing, all the exotic charged and neutral fermions, except for $N_0^5$ and $\nu^{0L}_4$, have Yukawa couplings
with scalars which get large vevs corresponding to the first stage of spontaneous symmetry breaking breaking. The new $N_0^5$ and 
$\nu^{0L}_4$ fields also need to get large masses, at least in GeV range, which can arise as discussed in the following sections.

 \subsection{Electroweak Breaking}
 \label{sec:electroweak-breaking} 

  Turning now to electroweak symmetry breaking, the corresponding vevs of the scalar fields are given as\\[-.5cm]
\begin{equation}\begin{split}
\langle S_c\rangle&=\begin{pmatrix}	0&0&0\\0& \langle (S_c)_{22} \rangle &0\\0&0&0\\	\end{pmatrix},\,
(\mathrm{U(1)}_\mathrm{X}, \mathbb{Z}_3, \mathbb{Z}_2) = (\frac 4 3, \omega^2, 1) \\
	\langle \eta\rangle&=\frac{1}{\sqrt 2}\begin{pmatrix} \langle \eta_1 \rangle \\0\\ \langle \eta_3 \rangle\end{pmatrix},\,
(\mathrm{U(1)}_\mathrm{X}, \mathbb{Z}_3, \mathbb{Z}_2) = (-\frac 1 3, \omega, 1) \\
	\langle \tilde{\eta} \rangle &=\frac{1}{\sqrt 2}\begin{pmatrix} \langle \tilde{\eta}_1 \rangle \\0\\ \langle \tilde{\eta}_3 \rangle \end{pmatrix},\,
(\mathrm{U(1)}_\mathrm{X}, \mathbb{Z}_3, \mathbb{Z}_2) = (-\frac 1 3, \omega, -1) \\
	\langle \rho\rangle&=\frac{1}{\sqrt 2}\begin{pmatrix}0\\ \langle \rho_2 \rangle \\0\end{pmatrix},\,
(\mathrm{U(1)}_\mathrm{X}, \mathbb{Z}_3, \mathbb{Z}_2) = (\frac 2 3, \omega^2, 1) \\
	\langle \tilde{\rho} \rangle&=\frac{1}{\sqrt 2}\begin{pmatrix}0\\ \langle \tilde{\rho}_2 \rangle \\0\end{pmatrix},\,
(\mathrm{U(1)}_\mathrm{X}, \mathbb{Z}_3, \mathbb{Z}_2) = (\frac 2 3, 1, 1)
\end{split}
\end{equation}


The neutral component of $L_5$ gets mass through invariant terms built with sextet, i.e.
\begin{equation}\begin{split}
&\bar L_5 S_c(L_5)^c\\
\end{split}
\end{equation}
This Yukawa term gives a diagonal mass term mass for the neutral  $N^0_5$. Note that since $S_c$ gets vev in its 22-component, a large value of $\vev{ (S_c)_{22}}$ will change the
  $\rho$-parameter from its canonical SM value. Therefore, the vev of $S_c$ field need to be small, less than 2 GeV or so. Thus, the dominant contribution to $N^0_5$ field's mass does
  not come from the above term but rather through its coupling with other fields (see~table~V), a fact that we have also checked numerically.

For the triplets, the relevant Yukawa terms for quarks and leptons are the following:\\[-1cm]
\begin{itemize}
\item 
	for quarks:
	\begin{equation}\begin{split}
	&\bar{q}_m \eta^* D\\
		&\bar{q}_3 \eta U\\
		&\bar{q}_3 \rho D\\
		&\bar{q}_m \rho^* U\end{split}
	\end{equation}
   where $D$ represents any right-handed $d_{1,2,3}$ or $B_{1,2}$,  $U$ represents any right-handed $u_{1,2,3}$, or $T_3$ and $m=1,2$

\item 
	for leptons:
		\begin{equation}\begin{split}
&\bar{\ell}_1 \eta^* e_{1,2}  \\
& \bar{\ell}_1 \tilde{\eta}^* E_{1}   \\
&\bar{\ell}_m \tilde{\rho} e_{1,2} \, \qquad \qquad  \qquad m = 2,3\\
& \bar{L}_4 \tilde{\rho} E_1 \\	
&\epsilon_{ijk} \tilde{\eta}^{*i}\bar\ell^{j}_m(L_5)^{c\,k} \, \qquad \; \; m = 2,3\\
& \epsilon_{ijk} \eta^{*i}\bar L^{j}_4 (L_5)^{c\,k}
\end{split}
\end{equation}
where the  $i,j,k=1,2,3$ indices refer to SU(3)$_\text{L}$. All these terms provide mass to charged leptons. The last two terms also provide mixing among neutral exotic states as well as mixing among SM  and exotic ones.
However, since the $\eta$ and $\tilde{\eta}$ vevs of electroweak level, none of these terms lead to unacceptable large masses for any SM particles, a fact that can be seen
  from the explicit forms of charged  and neutral lepton mass matrices given in Tables~(IV) and~(V) respectively. We have also numerically cross-checked this fact.
\end{itemize}

Actually, another Higgs sextet $S_b$ would be allowed by the symmetries of the model, with vev as
\begin{equation}
\langle S_b\rangle = \begin{pmatrix}
\langle(S_b)_{11}\rangle & 0 & \langle(S_b)_{13}\rangle \\
0 & 0 & 0 \\
\langle(S_b)_{13}\rangle & 0 & \langle(S_b)_{33}\rangle 
\end{pmatrix},\,
(U(1)_X, \mathbb{Z}_3, \mathbb{Z}_2) = (-\frac{2}{3}, \omega^2, 1) \\  \nonumber
\end{equation}
leading to the $ U(1)_X \otimes \mathbb{Z}_3 \otimes \mathbb{Z}_2$ Majorana mass terms
\begin{equation}\begin{split}
&\bar\ell_nS_b(\ell_m)^c,\quad n,m=2,3\\
&\bar L_4 S_b (L_4)^c \nonumber
\end{split} 
\end{equation}
The first of these terms  gives rise to diagonal mass terms for left-handed neutrinos  of the order of the EW scale. Therefore, in order to get the  observed tiny neutrino masses through a seesaw mechanism, we  exclude  the $S_b$ sextet from the particle content.

\subsection{Type-I Seesaw mechanism in 331-setup}
\label{sec:type-i-seesaw}

 For implementing the Type-I seesaw mechanism we need the following terms
	\begin{equation}\begin{split}
&    \bar{\ell}_m \, \eta \, \nu^R_a   \\
&\bar{L}_4 \, \tilde{\eta} \, \nu^R_a  \\
&  \bar{\nu}^R_a (\nu_b^{R})^c
  \label{seesaw-terms1-z2}
  \end{split}
\end{equation}
where $m = 2,3$ and 
$a,b = 1,2,3$.  They provide Dirac and Majorana masses for the SM-like neutrinos as well as their mixing with heavy neutral fermions.
The second term in \eqref{seesaw-terms1-z2} differs from the first, since $\ell_m$ is replaced by $L_4$.
They are distinct thanks to the $\mathbb{Z}_2$ symmetry. This ensures that the neutrino-like fermion in $L_4$ receives an adequately large mass because of suitably tuned Yukawa coupling. 

In  addition the following terms are also allowed by all the symmetries of the model
 \begin{equation}\begin{split}
 &\bar{\ell}_m \, \chi \, \nu^R_a \\
& \bar{L}_4 \, \tilde{\chi} \, \nu^R_a  \\
&
  \bar{\ell}_1 \, \tilde{\rho}^* \, \nu^R_a  
   \label{seesaw-terms2-z2}
  \end{split}
\end{equation}
As in the previous case, the first two terms in \eqref{seesaw-terms2-z2} are distinct due to the $\mathbb{Z}_2$ symmetry
(though in this case, a single term would not be dangerous as it would only give mass to the third component of $\ell_m$  due to vev alignment of $\chi$).

\section{ Fermion Mass Matrices}
\label{sec:ferm-mass-matr}

In the  full Yukawa Lagrangian characterizing our model   %
\begin{itemize}
	\item for quarks we have
	\begin{equation}
	\begin{split}
		\mathcal{L}^{q}_Y&=\bigl(\bar q_m\chi^*Y^d_{mi}+\bar q_3\rho y^d_{3i}+\bar q_m \eta^* j^d_{mi}\bigr)D_{i}+\\&+\bigl(\bar q_3\chi Y^u_{3j}+\bar q_m\rho^*y^u_{mj}+\bar q_3\eta j^u_{3j}\bigr)U_{j},
	\end{split}
		\label{eq:yukq}
	\end{equation}
	where $Y^{d,u}, y^{d,u}, j^{d,u}$ represent the Yukawa couplings introduced respectively for $\chi, \rho$ and $\eta$.    We remind that $D$ represents any right-handed $d_{1,2,3}$ or $B_{1,2}$,  $U$ represents any right-handed $u_{1,2,3}$, or $T_3$, and $m=1,2$. 
      \item for leptons we have
        \begin{equation}
\label{eq:yuklep}
\begin{split}
\mathcal{L}^{\ell}_Y = &  \bigl(Y_{1a} \bar\ell_1\chi^* + f_{ma} \bar\ell_m\rho
+ y_{1a} \bar\ell_1\eta^* \bigr)e_{a}  \, + \,\bigl(Y_{1E} \bar\ell_1 \tilde{\chi}^* 
+ y_{1E} \bar\ell_1 \tilde{\eta}^* \bigr) E_1 \, + \, f_{4E} \bar L_4 \tilde{\rho} E_1   \\
&    \, + \,  J_m \epsilon_{ijk}(\tilde{\chi}^*)^i(L_5)^{c\,k} \bar\ell_m^{j} 
\, + \,  J_4 \epsilon_{ijk}(\chi^*)^i(L_5)^{c\,k} \bar{L}_4^{j}
\, + \, j_{m} \epsilon_{ijk}(\tilde{\eta}^*)^i(L_5)^{c\,k} \bar\ell_m^{j} 
\\
& \, + \, j_{4} \epsilon_{ijk}(\eta^*)^i(L_5)^{c\,k} \bar{L}_4^{j}  + \frac{K_{mn}}{\sqrt{2}}  \bar\ell_m S_1(\ell_n)^c \, + \, \frac{K_{44}}{\sqrt{2}}  \bar{L}_4 S_1 (L_4)^c  + \frac{c_5}{\sqrt{2}} \bar L_5S_c(L_5)^c+      
\\
&  \, + \, (y_\eta)_{m s} \bar{\ell}_m \eta \nu_s^R \, + \, (y_{\tilde{\eta}})_{4s} \bar{L}_4 \tilde{\eta} \nu_s^R  \,+ \, (Y_\chi)_{ms} \bar{\ell}_m \chi \nu_s^R  \,+ \, (Y_{\tilde{\chi}})_{4 s} \bar{L}_4 \tilde{\chi} \nu_s^R
\, + \, (y_{\tilde{\rho}})_{1s} \bar{\ell}_1 \tilde{\rho}^* \nu_s^R \\
& \,+ \, \frac{M^{s t}}{\sqrt{2}} \bar{\nu}_s^R (\nu_t^R)^c  \, + \, \text{h.c.}
	\end{split}
	\end{equation}
      where $Y, y,  K, k, f,  c, J, j, M$ represent the Yukawa couplings, with $m,n \in \{2,3\}$, $a,b \in \{1,2)$,  $s,t \in  \{1,2,3\}$, and the $i,j,k \in \{1,2,3\}$  indices act on SU(3)$_\text{L}$.
\end{itemize}

 The mass matrices for the up-type quarks ($\sqrt{2} M^u_{ij}$), shown in Table.~\ref{tab:up}, and down-type quarks ($\sqrt{2} M^d_{ij}$), shown in Table.\ref{tab:down}  remain exactly the same as before, namely 
  \begin{table}[h!t]
\centering
\begin{tabular}{| c | c | c | c | c |  }
  \hline 
 Fields    & \hspace{.05cm} $u^R_1$ \hspace{.05cm} & \hspace{.05cm} $u^R_2$ \hspace{.05cm} & \hspace{.05cm} $u^R_3$ \hspace{.05cm} & \hspace{.05cm} $T^R_3$ \hspace{.05cm}  \hspace{.05cm}    \\
\hline 
$\bar{u}^L_1$  &  $ - y^u_{11} \langle \rho^*_2\rangle $  & $ - y^u_{12} \langle \rho^*_2\rangle $  &   
$ - y^u_{13} \langle \rho^*_2\rangle $ &  $ - y^u_{14} \langle \rho^*_2\rangle $ 
\\ \hline
$\bar{u}^L_2$  &  $ - y^u_{21} \langle \rho^*_2\rangle $  & $ - y^u_{22} \langle \rho^*_2\rangle $  &   
$ - y^u_{23} \langle \rho^*_2\rangle $ &  $ - y^u_{24} \langle \rho^*_2\rangle $            
\\ \hline
$\bar{u}^L_3$  &  $ j^u_{31} \langle \eta_1 \rangle $   &  $ j^u_{32} \langle \eta_1 \rangle $    &  
 $ j^u_{33} \langle \eta_1 \rangle $  &   $ j^u_{34} \langle \eta_1 \rangle $            
\\ \hline
$\bar{T}^L_3$  &  $ Y^u_{31} \langle \chi_3 \rangle + j^u_{31} \langle \eta_3 \rangle$          &
$ Y^u_{32} \langle \chi_3 \rangle + j^u_{32} \langle \eta_3 \rangle$   &  $ Y^u_{33} \langle \chi_3 \rangle + j^u_{33} \langle \eta_3 \rangle$ & $ Y^u_{34} \langle \chi_3 \rangle + j^u_{34} \langle \eta_3 \rangle$       
\\ \hline
  \end{tabular}
  \caption{
   Up-type quark mass matrix $\sqrt{2} M^u_{ij}$. Here the $L$ and $R$ superscripts indicate the left and right-handed fields.}
   \label{tab:up}
\end{table}
 \begin{table}[h!t]
\centering
\begin{tabular}{| c | c | c | c | c | c | }
  \hline 
 Fields    & \hspace{.05cm} $d^R_1$ \hspace{.05cm} & \hspace{.05cm} $d^R_2$ \hspace{.05cm} & \hspace{.05cm} $d^R_3$ \hspace{.05cm} & \hspace{.05cm} $B^R_1$ \hspace{.05cm} & \hspace{.05cm} $B^R_2$ \hspace{.05cm}    \\
\hline 
$\bar{d}^L_1$  &  $ j^d_{11} \langle \eta^*_1\rangle $   &   $ j^d_{12}\langle \eta^*_1\rangle $  &   
$ j^d_{13}\langle \eta^*_1\rangle $  &  $ j^d_{14} \langle \eta^*_1 \rangle $  &  $ j^d_{15} \langle \eta^*_1 \rangle $ 
\\ \hline
$\bar{d}^L_2$  &  $ j^d_{21} \langle \eta^*_1\rangle $   &   $ j^d_{22}\langle \eta^*_1\rangle $  &   
$ j^d_{23}\langle \eta^*_1\rangle $  &  $ j^d_{24} \langle \eta^*_1 \rangle $  &  $ j^d_{25} \langle \eta^*_1 \rangle $    
\\ \hline
$\bar{d}^L_3$  &  $ y^d_{31} \langle \rho_2 \rangle $   &   $ y^d_{32}\langle \rho_2\rangle $  &   
$ y^d_{33}\langle \rho_2\rangle $  &  $ y^d_{34} \langle \rho_2 \rangle $  &  $ y^d_{35} \langle  \rho_2 \rangle $  
\\ \hline
$\bar{B}^L_1$  &  $ Y^d_{11} \langle \chi^*_3 \rangle + j^d_{11} \langle \eta^*_3 \rangle$          &
$ Y^d_{12} \langle \chi^*_3 \rangle + j^d_{12} \langle \eta^*_3 \rangle$   &  $ Y^d_{13} \langle \chi^*_3 \rangle + j^d_{13} \langle \eta^*_3 \rangle$   & $ Y^d_{14} \langle \chi^*_3 \rangle + j^d_{14} \langle \eta^*_3 \rangle$   & $ Y^d_{15} \langle \chi^*_3 \rangle + j^d_{15} \langle \eta^*_3 \rangle$    
\\ \hline
$\bar{B}^L_2$  &  $ Y^d_{21} \langle \chi^*_3 \rangle + j^d_{21} \langle \eta^*_3 \rangle$          &
$ Y^d_{22} \langle \chi^*_3 \rangle + j^d_{22} \langle \eta^*_3 \rangle$   &  $ Y^d_{23} \langle \chi^*_3 \rangle + j^d_{23} \langle \eta^*_3 \rangle$   & $ Y^d_{24} \langle \chi^*_3 \rangle + j^d_{24} \langle \eta^*_3 \rangle$   & $ Y^d_{25} \langle \chi^*_3 \rangle + j^d_{25} \langle \eta^*_3 \rangle$      
\\    \hline
  \end{tabular}
  \caption{
  Down-type mass matrix $\sqrt{2} M^d_{ij}$. Here the $L$ and $R$ superscripts indicate the left and right-handed fields.}
 \label{tab:down}
\end{table}

  Turning to the lepton mass matrices, we begin with charged lepton mass matrix ($\sqrt{2} M^e_{ij}$), whose explicit form is given in Table~\ref{tab:chlep}.
\begin{table}[h!t]
\centering
\begin{tabular}{| c | c | c | c | c | c | }
  \hline 
 Fields    & \hspace{.05cm} $e^R_1$ \hspace{.05cm} & \hspace{.05cm} $e^R_2$ \hspace{.05cm} & \hspace{.05cm} $e^R_3$ \hspace{.05cm} & \hspace{.05cm} $E^R_1$ \hspace{.05cm} & \hspace{.05cm} $E^R_4$ \hspace{.05cm}    \\
\hline 
$\bar{e}^L_1$  &  $ y_{11} \langle \eta^*_1\rangle $   &   $ y_{12}\langle \eta^*_1\rangle $  &   
$0$            &  $ y_{1E} \langle  \tilde{\eta}^*_1 \rangle $  &  $0 $ 
\\ \hline
$\bar{e}^L_2$  &  $ f_{21} \langle \rho_2 \rangle $       &   $ f_{22} \langle \rho_2 \rangle $     &  
$ j_2 \langle \tilde{\eta}^*_1 \rangle $  &  $0$            
& $ -(J_2 \langle \tilde{\chi}^*_3 \rangle +  j_2 \langle \tilde{\eta}^*_3 \rangle)$   
\\ \hline
$\bar{e}^L_3$  &  $ f_{31} \langle \rho_2 \rangle $       &  $ f_{33} \langle \rho_2 \rangle $     &  
$ j_3 \langle \tilde{\eta}^*_1 \rangle $  &  $0$          
& $ -(J_3 \langle \tilde{\chi}^*_3 \rangle +  j_3 \langle \tilde{\eta}^*_3 \rangle)$  
\\ \hline
$\bar{E}^L_1$  &  $ Y_{11} \langle \chi^*_3 \rangle + y_{11} \langle \eta^*_3 \rangle$          &
$ Y_{12} \langle \chi^*_3 \rangle + y_{12} \langle \eta^*_3 \rangle$   &  $ 0 $                 & 
$ Y_{1E} \langle \tilde{\chi}^*_3 \rangle + y_{1E} \langle \tilde{\eta}^*_3 \rangle$   & $ 0 $    
\\ \hline
$\bar{E}^L_4$  &  $0 $ & $ 0 $ & $ j_4 \langle \eta^*_1 \rangle $  & $ f_{4E} \langle \tilde{\rho}_2 \rangle $         & $ -(J_4 \langle \chi^*_3 \rangle + j_4 \langle \eta^*_3 \rangle) $   
\\    \hline
  \end{tabular}
  \caption{
      The charged lepton mass matrix $\sqrt{2} M^e_{ij}$. Here subscripts of the vev-carrying scalars indicate the scalar compenents whose non-zero vev comes in a given entry.   
          }
 \label{tab:chlep}
\end{table}

Concerning the mass matrix of the neutral fermions ($\sqrt{2} M^n_{ij}$), it incorporates type-I seesaw mass terms.
  Its complete form is given in the Appendix, Table V.
We have numerically verified that it leads to an adequate spectrum of light neutrino masses.

\section{B flavour global analyses}
\label{sec:b-flavour-global}

These analyses are performed in the framework of the effective Hamiltonian at the $b$-mass scale, separating short- and long-distance physics in the Wilson coefficients and local operators~\cite{Grinstein:1987vj, Buchalla:1995vs}:
\begin{equation}
  {\mathcal H}_{\rm eff}=-\frac{4G_F}{\sqrt{2}} V_{tb} V_{ts}^* \sum_i C_i O_i
\end{equation}
 The main operators of interest for this discussion are the following:
\begin{equation}
\begin{split}
        O_7=&\frac{e}{16\pi^2} m_b (\bar s \sigma_{\mu\nu} P_R b)F^{\mu\nu}\\
        O_{7'}=&\frac{e}{16\pi^2} m_b (\bar s \sigma_{\mu\nu} P_L b)F^{\mu\nu}\\
	O_9^\ell=&\frac{e^2}{16\pi^2}(\bar s \gamma_{\mu} P_L b) (\bar\ell\gamma^\mu \ell)\\
	O_{10}^\ell=&\frac{e^2}{16\pi^2}(\bar s \gamma_{\mu} P_L b) (\bar\ell\gamma^\mu\gamma^5 \ell)\\
	O_{9'}^\ell=&\frac{e^2}{16\pi^2}(\bar s \gamma_{\mu} P_R b) (\bar\ell\gamma^\mu \ell)\\
	O_{10'}^\ell=&\frac{e^2}{16\pi^2}(\bar s \gamma_{\mu} P_R b) (\bar\ell\gamma^\mu\gamma^5 \ell).\\
\end{split}
\label{eq:OP}
\end{equation}
where $P_{L,R}=(1\mp \gamma_5)/2$ and the fields are understood as mass eigenstates.
In the SM, only  $O_7$, $O_9^\ell$ and $O_{10}^\ell$ are significant, with the values of the Wilson coefficients given as $C_9^\ell\simeq 4.1$ and
$C_{10}^\ell\simeq -4.3$ at the scale $\mu=m_b$. In contrast, the primed operators are $m_s/m_b$ suppressed due to the chirality of the quarks involved. 

The analyses of several $b\to s\gamma$ and $b\to s\ell\ell$ observables (including angular ones) point towards
a pattern of deviations consistent with a large NP short-distance contribution to $C_9^\mu$,
around 1/4 of the SM contribution, see e.g. Refs.~\cite{Descotes-Genon:2015uva, Descotes-Genon:2016hem,Capdevila:2017bsm, Hiller:2003js}.  
Scenarios with NP contributions in $C_9^\mu$ only, in $(C_9^\mu,C_{10}^\mu)$ or  in $(C_9^\mu,C_{9'}^\mu)$ seem particularly favoured.  
Moreover, the LFU violating observables agree well with the absence of significant NP contributions to any electron-type Wilson coefficients $C_{i}^{e}$.  
Results of the global fit analyses seem to rule out the possibility of large contributions from other operators suppressed in the SM, in particular scalar and pseudoscalar operators. They are constrained especially by the good agreement between the observed value for the $B_s \to \mu \mu$ branching ratio and its SM prediction, as well as by the limits on the $B \to X_s \gamma$ branching ratio.

We  proceed along the lines of  the phenomenological analysis of Ref. \cite{Descotes-Genon:2017ptp} to which we refer for details.
We focus on the vector/axial contributions which are assumed to be the larger ones.
The neutral lepton mass matrix and the neutral lepton mixing do not affect the effective Hamiltonian contributing to the process, since the relevant operators only include  charged leptons.
Hence after the expansion in $\epsilon=\Lambda_\text{EW}/\Lambda_\text{NP}$ {(NP denotes here the 331 scale) one finds that nonzero contributions at the lowest order,
namely $O(\epsilon^2)$, can only come from the neutral gauge bosons $Z'$ and $Z$. 
The transitions mediated by the heavy gauge boson $Z'$ are expressed in the effective Hamiltonian by the term 
{%
\begin{widetext}\begin{eqnarray}\label{eff:Z'}
	\mathcal H_{\text{eff}}&\supset &\frac{g_X^2}{54\cos^2\theta_{331}}\frac{1}{M^2_{Z'}}V^{(d)*}_{3k}V^{(d)}_{3l}
	\frac{4\pi}{\alpha}
	\\\nonumber
	&&	
	\Biggl\{\left[-\frac 1 2 V^{(e)*}_{1i}V^{(e)}_{1j}+\frac{1-6\cos^2\theta_{331}}2W^{(e)*}_{3i}W^{(e)}_{3j}+\frac{1+3\cos^2\theta_{331}}{4}\delta_{ij}\right]O^{klij}_9+\\
	&&\qquad+\left[\frac 1 2 V^{(e)*}_{1i}V^{(e)}_{1j}+\frac{1-6\cos^2\theta_{331}}2W^{(e)*}_{3i}W^{(e)}_{3j}+\frac{-1+9\cos^2\theta_{331}}{4}\delta_{ij}\right]O^{klij}_{10}\Biggr\}.\nonumber
\end{eqnarray}\end{widetext}
where the indices $k,l$ refer to the SM generations of the quark mass eigenstates (assuming $k\neq l$),  while $i, j$ refer to the SM lepton mass eigenstates
(either from the same or different generations).
The effective operators $O_{{9,10}}^{klij}$ are defined exactly as in Eq.~\eqref{eq:OP}, taking into account  the $(\bar q_k\, q_l) (\bar \ell_i\, \ell_j)$ flavour structure.
Here $\alpha=e^2/(4 \pi)$ is the fine-structure constant.
The $V$ and $W$ matrices provide the mixing matrices arising from the diagonalisation of the EWSB mass terms in the subspace of left-handed and right-handed SM fields,
with the superscript $(d)$ and $(e)$ referring to down-type quarks and charged leptons, respectively. 

At the same lowest order the contribution to the effective Hamiltonian given by the SM gauge boson $Z$ can be written as
\begin{equation}\begin{split}
	\mathcal H_{\text{eff}}\supset \frac{\cos^2\theta_W(1+3\cos^2\theta_{331})}{8}\frac{g^2}{M^2_{Z}}\frac{4\pi}{\alpha}\sum_\lambda \hat{V}^{(d)*}_{\lambda k}\hat{V}^{(d)}_{\lambda l}\delta_{ij}\times\\\times\Bigl\{(-1+9\cos^2\theta_{331})O^{klij}_9+(1+3\cos^2\theta_{331})O^{klij}_{10}\Bigr\}.
\label{eff:Z}\end{split}
\end{equation}
 where $\hat{V}^{(d)}$ represents the $O(\epsilon^1)$ correction to the rotation matrix $V^{(d)}$ between interaction and mass eigenstates for the left-handed down sector. 
 Notice that at this order the coupling is the same for all the light leptons, i.e. non-universality does not arise  in the interaction with $Z$.
 LFU violating contributions arise only from the $Z'$ contribution. 
 
 In addition to LFU violation, the model allows for lepton-flavour violation, which we assume suppressed, in agreement with experimental restrictions, and set it to zero for simplicity.
 These further assumptions constrain the parameter space $(C_9^\mu,C_{10}^\mu)$ to two scenarios detailed in Ref. \cite{Descotes-Genon:2017ptp}.
 For both of them, we can compare the allowed regions with the latest data, as done in Fig.~\ref{fig:caseA}. In this figure, and from now on,
 we focus only on the non SM contribution to the Wilson coefficients, that is we set $C_i=C_i^{NP}$.
 The thick black intervals correspond to the 1$\sigma$ interval for the one-dimensional scenarios from the latest data \cite{Alguero:2021anc}.
\begin{figure}[h]
\centering
\includegraphics[width=0.3\textwidth]{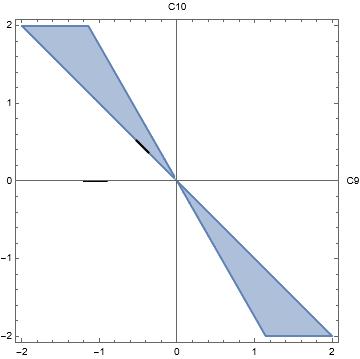}
\includegraphics[width=0.3\textwidth]{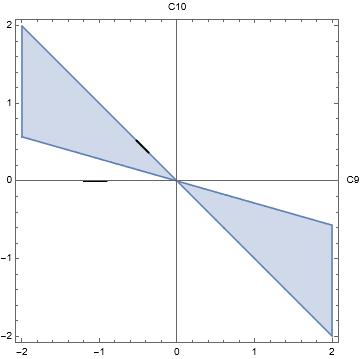}
\caption{Regions allowed for the Wilson coefficient $C_9^{\mu}$ and $C_{10}^{\mu}$ (abscissa and ordinate, respectively) in scenarios A (left) and B (right) described in Ref. 
\cite{Descotes-Genon:2017ptp}. The thick black intervals correspond to the 1$\sigma$  interval for one-dimensional scenarios \cite{Alguero:2021anc}. }
\label{fig:caseA}
\end{figure}

A comparison between 2018 and 2021 intervals for $C_{9\mu}$ given by global analyses \cite{Capdevila:2017bsm, Alguero:2021anc} is reported below:
\begin{itemize}
\item $C_{9\mu}$, $C_{10\mu}=0$
\begin{equation}
[-1.28,-0.94], \quad (2018)
\end{equation} 
\begin{equation}
[-1.20,-0.91], \quad (2021) 
\end{equation}
\item $C_{9\mu}=-C_{10\mu}$
\begin{equation}
[-0.75,-0.49], \quad (2018) 
\end{equation} 
\begin{equation}
[-0.52,-0.37], \quad (2021) 
\end{equation}
\end{itemize}
%
As can be seen in Fig.~\ref{fig:caseA}, also with new data in both scenarios A and B we are able to account for the anomalies observed as long as we consider the $C_9^\mu=-C_{10}^\mu$  case.

In our model the $b \to s \ell \ell$ transitions originate from the tree-level exchange of the $Z$ and $Z'$ gauge bosons.
  The former breaks the GIM mechanism through the mixing between normal and exotic quarks, and depends on the Yukawa couplings.
  The latter involves just the unsuppressed exchange of the heavy $Z'$ gauge boson.
  Both give suppressed contributions to the $bsZ$ vertex, as can be seen on Fig. \ref{fig:Bsmix}.
To make a quantitative analysis we must take into account phenomenological constraints on $Z$ and $Z'$ couplings.

Restricting our discussion to the leading contributions of order ${\mathcal O}(\epsilon^2)$,
the $Z$-exchange contribution to $B_s-\bar{B}_s$ mixing will have two such vertices, and hence the amplitude will be suppressed by a factor ${\mathcal O}(\epsilon^4)$.
On the other hand, the $bs$ vertex is mediated by $Z'$ at ${\mathcal O}(\epsilon^0)$, implying that in this case we have only the suppression coming from
the heavy propagator must be taken into account.
\begin{figure}[h]
\centering
\includegraphics[width=0.5\textwidth]{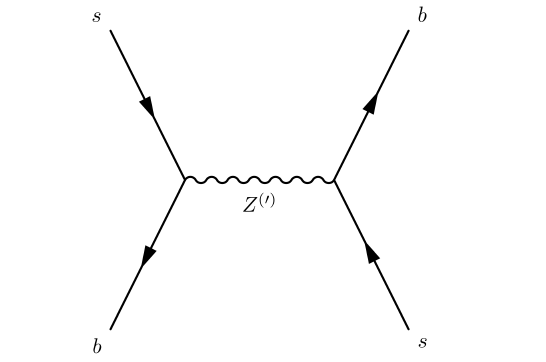}
\caption{Tree level contributions to $B_s-\bar{B}_s $ mixing.}
\label{fig:Bsmix}
\end{figure}
The corresponding part of the effective Hamiltonian  is
\begin{equation}
	\begin{split}
		&\mathcal H_{\rm eff}\supset \frac{g_X^2}{54M^2_{Z'}\cos^2 \theta_{331}}(V^{*(d)}_{3k}V^{(d)}_{3l})^2(\overline{D_k}\gamma^\mu D_l)(\overline{D_k}\gamma^\mu D_l)=\\
					&=\frac{8G_F}{\sqrt 2 (3-\tan^2\theta_W)}\frac{M_W^2}{M_{Z'}^2}(V^{*(d)}_{3k}V^{(d)}_{3l})^2(\overline{D_k}\gamma^\mu D_l)(\overline{D_k}\gamma^\mu D_l)\\
	\end{split}
\end{equation}
Our case of interest is  $k=2,l=3$.\
The SM contribution to the mixing reads~\cite{Lenz:2010gu}
\begin{equation}
	\mathcal H^\text{SM}_{\rm eff} = (V_{ts}^*V_{tb})^2\frac{G_F^2}{4\pi^2} M_W^2 \hat{\eta}_B S\Bigl(\frac{\overline{m_t}^2}{M_W^2}\Bigr)(\overline{s_L}\gamma^\mu b_L)(\overline{s_L}\gamma^\mu b_L)
\end{equation}
where $S$ is the Inami-Lim function and $\overline{m_t}$ is  the top quark mass defined in the
$\overline{MS}$ scheme. As in Ref.~\cite{Lenz:2010gu}, we take  $S\Bigl(\frac{\overline{m_t}^2}{M_W^2}\Bigr)\simeq 2.35$, for a top mass of about 165 GeV, and $\hat{\eta}_B=0.8393\pm 0.0034$,
which includes QCD corrections. Considering the modulus of the ratio of the NP contribution over the SM, one gets
\begin{equation}\begin{split}
	r_{B_s}&=\left|\frac{C_\text{NP}}{C_\text{SM}}\right| =\\&= \frac{32\pi^2|V^{*(d)}_{32}V^{(d)}_{33}|^2}{\sqrt 2 (3-\tan^2\theta_W)|V_{ts}^*V_{tb}|^2G_FM_W^2\hat{\eta}_B S} \frac{M_W^2}{M_{Z'}^2}\end{split}
\end{equation}
Here the only variables are $d=V^{*(d)}_{32}V^{(d)}_{33}$ and $M_{Z'}^2$ or, equivalently, $M_W^2/M_{Z'}^2$. In order to get a quantitative idea of the values allowed,
we perform a scan varying $d$ in $[-1,1]$ (since $d$ consist of products of elements of unitary matrices).
We fix the range of the other variable  $M_W/M_{Z'}$ to $[0,0.1]$, corresponding roughly to a NP scale at
least of the order of 10 times the electroweak scale, and assume that   the NP contributions to the $B_s$ mixing is at most $10\%$ by setting $r_{B_s}\leq 0.1$.
For those values, we evaluate the NP contribution to the Wilson coefficient in the one-dimensional scenario with $C_9^\mu=-C_{10}^\mu$.
The allowed values found in the scan are plotted in Fig.~\ref{fig:scanBB}.
%
\begin{figure}[h]
\centering
\includegraphics[width=0.5\textwidth]{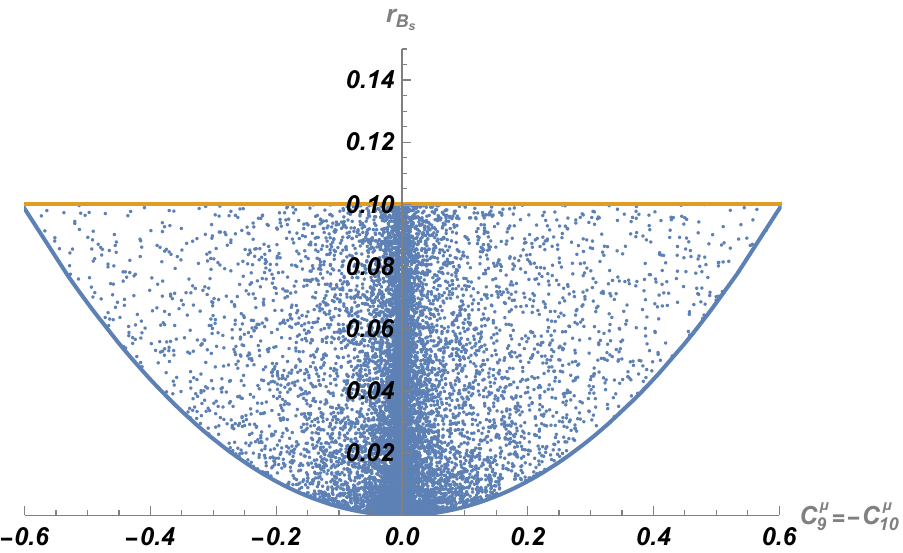}
\caption{Allowed points in the ($C^\mu_{9},r_{B_s}$) plane.}
\label{fig:scanBB}
\end{figure}

We see that values of $C_9^\mu=-C_{10}^\mu$ can reach -0.6, in agreement with the results of global analyses of $b\to s\ell\ell$, corresponding to $r_{B_s}=0.1$, $M_W/M_{Z'}=0.1$ and $d\simeq  -0.005$.
The allowed region is limited by the fact that we have numerically taken
\begin{equation}\begin{split}
r_{B_s}&\simeq 347\cdot 10^3 \times  \left(\frac{M_W}{M_{Z'}}\right)^2 \times d^2\leq 0.1\\
C_9^\mu &\simeq 11.3 \cdot 10^3 \times \left(\frac{M_W}{M_{Z'}}\right)^2 \times d\qquad |d|\leq 1\end{split}
\end{equation}
Therefore, in the simple one-dimensional scenario $C_9^\mu=-C_{10}^\mu$, the present 331 model can accomodate both $B_s-\bar{B}_s$ mixing and $b \to s \ell \ell$ data, with a NP scale
(and in particular a $Z'$) around the TeV scale.
Searches for high-mass dilepton resonances at ATLAS~\cite{ATLAS:2019erb} have set higher lower limits for $Z'$ by comparison with different 331 models~\cite{Queiroz:2016gif}.
As the limits on the $Z'$ mass from direct searches gets higher, our points are pushed towards the plot edges, requiring a larger value of $r_{B_s}$.
However, care must be used to extrapolate results from other 331 models, especially minimal ones, since different couplings and interference patterns may affect the results of the searches.
The lower bounds of $Z'$  mass can be significantly lower than those obtained from LHC, if all decay channels of $Z'$ into new particles are included.

\section{Summary and outlook} 
\label{Conclusion}

In this paper we have explored the possibility of explaining data on flavour anomalies for $B \to  K^{(*)}$ decays within a 331 extension of the Standard Model.
We explored the possibility of having a new massive 331 $Z'$ boson coupled in a different way to muons and electrons. 
We are aware of the intrinsic limitations of fiddling with gauge couplings in the absence of a dedicated family symmetry.
Nevertheless our analysis is encouraged by previous results in Ref. \cite{Descotes-Genon:2017ptp},
and motivated by recent data that tend to confirm flavour anomalies; in particular, 2021 data of LHCb achieve 
a $3.1\sigma$ deviation from SM predictions in the $R_{K^{(*)}} $ observable in $B^+ \to  K^+ \ell^+ \ell^-$ decays
with 9 fb$^{-1}$ of proton-proton collision data \cite{LHCb:2021trn}.

Prompted by these new data we examine the viability of generalizing the scheme in Ref. \cite{Descotes-Genon:2017ptp} so as to provide
a complete 331 model explaining LFU violation and generating viable neutrino masses through a type-I seesaw mechanism.
We have shown the viability of a 331 gauge symmetry model setup putting together both flavour anomalies and a consistent neutrino mass spectrum. 
The model introduces new massive particles at mass scales allowed by current laboratory data and requires a sophisticated structure beyond the "traditional" 331 schemes. 
Indeed, in order to eliminate dangerous mass terms and mixings 
our model employs a $\mathrm{SU(3)_{c}\times SU_{L}(3)\times U(1) } \times \mathbb{Z}_2 \times \mathbb{Z}_3$ symmetry. 
The new global discrete symmetries ensure a realistic mass hierarchy pattern for the fermions. 

Within the model-independent effective approach, deviations from lepton flavour universality in the $b \to s \ell \ell$ transitions are parameterized 
by new physics contributions to the Wilson coefficients.
Our extended 331 model can generate such large new physics contributions to $(C_9^\mu, C_{10}^\mu)$ parameters,
as required by current global fits \cite{Capdevila:2017bsm,Alguero:2021anc}. 
Trying to stick to minimality requirements, we assumed that neutral gauge bosons give a dominant contributions to the flavour violating observables
without contributions to $b \to s e e$ or large lepton flavour violation of the form $b\to s\ell_1\ell_2$, as suggested by experimental observations.
Within a simple one-dimensional scenario with opposite contributions to $C_9^\mu$ and $ C_{10}^\mu$,
we accommodate both $B_s-\bar{B}_s$ mixing and $b\to s\ell\ell$ data, with a new physics $Z'$ mass scale around the TeV scale.
Going to different values for $(C_9^\mu, C_{10}^\mu)$ would possibly extend the allowed parameter space for new physics. 
 In order to comply with experimental limits for processes involving charged leptons, we assume that contributions to $b \to s l_1 l_2$ as well as lepton-universality violating
 processes are suppressed. This allows us to set constraints on the fermionic mixing matrices, as discussed in Ref.~\cite{Descotes-Genon:2017ptp}.

In summary, we have reconciled the LFU violation data with a viable neutrino oscillation pattern in a 331 setup, a goal never achieved earlier.
Our explanation for $B$-anomaly decays may be reformulated within alternative neutrino mass generation mechanisms such as inverse seesaw mechanism.
Likewise, the inclusion of dark matter may be implemented through a scotogenic approach.

\begin{table}[p] \vspace{-1.5cm}
\centering\begin{tiny} 
\rotatebox{-90}{
\begin{minipage}{\textheight}
\begin{tabular}{| c | c | c | c | c | c | c | c | c | c | c | c | }
  \hline 
Fields    & \hspace{.05cm} $(\nu^L_1)^c$ \hspace{.05cm} & \hspace{.05cm} $(\nu^L_2)^c$ \hspace{.05cm} & \hspace{.05cm} $(\nu^L_3)^c$ \hspace{.05cm} & \hspace{.05cm} $(\nu^L_4)^c$ \hspace{.05cm} & \hspace{.05cm} $(N^L_2)^c$ \hspace{.05cm} & \hspace{.05cm} $(N^L_3)^c$ \hspace{.05cm} & \hspace{.05cm} $(N^L_4)^c$ \hspace{.05cm} & \hspace{.05cm} $(N^L_5)^c$ \hspace{.05cm} & \hspace{.05cm} $\nu^R_1$ \hspace{.05cm} & \hspace{.05cm} $\nu^R_2$ \hspace{.05cm} & \hspace{.05cm} $\nu^R_3$ \hspace{.05cm} 
\\ \hline 
$\bar{\nu}^L_1$      &  $0$  &  $0$   & $0$   & $0$   &  $0$  & $0$  &  $0$  &  $0$  &   
$-(y_{\tilde{\rho}})_{11}\langle \tilde{\rho}_2^*\rangle$  & $-(y_{\tilde{\rho}})_{12}\langle \tilde{\rho}_2^*\rangle$  &  $ -(y_{\tilde{\rho}})_{13}\langle \tilde{\rho}_2^*\rangle$
\\ \hline 
$\bar{\nu}^L_2$      &  $0$  &  $0$   & $0$   & $0$   &  $0$  &  $0$ &  $0$  &  $ J_2 \langle \tilde{\chi}^*_3\rangle + j_2 \langle \tilde{\eta}^*_3\rangle$  & $ (y_\eta)_{21} \langle \eta_1\rangle$  &  $ (y_\eta)_{22} \langle \eta_1\rangle$ &  $ (y_\eta)_{23} \langle \eta_1\rangle$ 
\\ \hline
$\bar{\nu}^L_3$      &  $0$  &  $0$   & $0$   & $0$   & $0$   & $0$  &  $0$  &  $ J_3 \langle \tilde{\chi}^*_3\rangle + j_3 \langle \tilde{\eta}^*_3\rangle$  & $ (y_\eta)_{31} \langle \eta_1\rangle$  &  $ (y_\eta)_{32} \langle \eta_1\rangle$ &  $ (y_\eta)_{33} \langle \eta_1\rangle$    
\\ \hline
$\bar{\nu}^L_4$      &  $0$  &  $0$   & $0$   & $0$   &  $0$  & $0$  &  $0$  &  $ J_4 \langle \chi^*_3\rangle + j_4 \langle \eta^*_3\rangle$  & $ (y_{\tilde{\eta}})_{41} \langle \tilde{\eta}_1\rangle$  &  $ (y_{\tilde{\eta}})_{42} \langle \tilde{\eta}_1\rangle$ &  $ (y_{\tilde{\eta}})_{43} \langle \tilde{\eta}_1\rangle$       
\\ \hline
$\bar{N}^L_2$        &  $0$  &  $0$   & $0$   & $0$   &  $K_{22} \langle S_1 \rangle$  &  $K_{23} \langle S_1 \rangle$ &  $0$  & $-j_2 \langle \tilde{\eta}^*_1 \rangle$   & $(y_\eta)_{21} \langle  \eta_3\rangle + (Y_\chi)_{21}\langle  \chi_3\rangle$  & $(y_\eta)_{22} \langle  \eta_3\rangle + (Y_\chi)_{22}\langle  \chi_3\rangle$  & $(y_\eta)_{23} \langle  \eta_3\rangle + (Y_\chi)_{23}\langle  \chi_3\rangle$  
\\ \hline  
$\bar{N}^L_3$        &  $0$  &  $0$   & $0$   & $0$   &   $K_{32} \langle S_1 \rangle$  &  $K_{33} \langle S_1 \rangle$ &  $0$  & $-j_3 \langle \tilde{\eta}^*_1 \rangle$    & $(y_\eta)_{31} \langle  \eta_3\rangle + (Y_\chi)_{31}\langle  \chi_3\rangle$  & $(y_\eta)_{32} \langle  \eta_3\rangle + (Y_\chi)_{32}\langle  \chi_3\rangle$  & $(y_\eta)_{33} \langle  \eta_3\rangle + (Y_\chi)_{33}\langle  \chi_3\rangle$    
\\ \hline
$\bar{N}^L_4$        &  $0$  &  $0$   & $0$   & $0$   &  $0$  &  $0$ & $K_{44} \langle S_1 \rangle$   &  $-j_4 \langle \eta^*_1 \rangle$   & $(y_{\tilde{\eta}})_{41} \langle  \tilde{\eta}_3\rangle + (Y_{\tilde{\chi}})_{41}\langle \tilde{\chi}_3\rangle$  & $(y_{\tilde{\eta}})_{42} \langle  \tilde{\eta}_3\rangle + (Y_{\tilde{\chi}})_{42}\langle \tilde{\chi}_3\rangle$   & $(y_{\tilde{\eta}})_{43} \langle  \tilde{\eta}_3\rangle + (Y_{\tilde{\chi}})_{43}\langle \tilde{\chi}_3\rangle$     
\\ \hline
$\bar{N}^L_5$        &  $0$  &  $ J_2 \langle \tilde{\chi}_3\rangle + j_2 \langle \tilde{\eta}_3\rangle$    & $ J_3 \langle \tilde{\chi}_3\rangle + j_3 \langle \tilde{\eta}_3\rangle$   & $ J_4 \langle \chi_3\rangle + j_4 \langle \eta_3\rangle$   &  $-j_2 \langle \tilde{\eta}_1 \rangle$  &   $-j_3 \langle \tilde{\eta}_1 \rangle$ &  $-j_4 \langle \eta_1 \rangle$  & $c_5 \langle S_c \rangle$   &  $0$  &  $0$  &   $0$ 
\\ \hline
$(\bar{\nu}^R_1)^c$  & $-(y_{\tilde{\rho}})_{11}\langle \tilde{\rho}_2\rangle$   & $ (y_\eta)_{21} \langle \eta_1^*\rangle$  & $ (y_\eta)_{31} \langle \eta_1^*\rangle$  & $ (y_{\tilde{\eta}})_{41} \langle \tilde{\eta}_1^*\rangle$     &  $(y_\eta)_{21} \langle  \eta_3^*\rangle + (Y_\chi)_{21}\langle  \chi_3^*\rangle$   & $(y_\eta)_{31} \langle  \eta_3^*\rangle + (Y_\chi)_{31}\langle  \chi_3^*\rangle$  & $(y_{\tilde{\eta}})_{41} \langle  \tilde{\eta}_3^*\rangle + (Y_{\tilde{\chi}})_{41}\langle \tilde{\chi}_3^*\rangle$   &  $0$  &  $M_{11}$  & $M_{12}$  &  $M_{13}$    
\\ \hline
$(\bar{\nu}^R_2)^c$  &  $-(y_{\tilde{\rho}})_{12}\langle \tilde{\rho}_2\rangle$   & $ (y_\eta)_{22} \langle \eta_1^*\rangle$  & $ (y_\eta)_{32} \langle \eta_1^*\rangle$  & $ (y_{\tilde{\eta}})_{42} \langle \tilde{\eta}_1^*\rangle$     &  $(y_\eta)_{22} \langle  \eta_3^*\rangle + (Y_\chi)_{22}\langle  \chi_3^*\rangle$   & $(y_\eta)_{32} \langle  \eta_3^*\rangle + (Y_\chi)_{32}\langle  \chi_3^*\rangle$  & $(y_{\tilde{\eta}})_{42} \langle  \tilde{\eta}_3^*\rangle + (Y_{\tilde{\chi}})_{42}\langle \tilde{\chi}_3^*\rangle$    &  $0$ &  $M_{21}$ &  $M_{22}$  &  $M_{23}$
\\ \hline
$(\bar{\nu}^R_3)^c$  & $ -(y_{\tilde{\rho}})_{13}\langle \tilde{\rho}_2\rangle$   & $ (y_\eta)_{23} \langle \eta_1^*\rangle$  & $ (y_\eta)_{33} \langle \eta_1^*\rangle$  & $ (y_{\tilde{\eta}})_{43} \langle \tilde{\eta}_1^*\rangle$     &  $(y_\eta)_{23} \langle  \eta_3^*\rangle + (Y_\chi)_{23}\langle  \chi_3^*\rangle$   & $(y_\eta)_{33} \langle  \eta_3^*\rangle + (Y_\chi)_{33}\langle  \chi_3^*\rangle$  & $(y_{\tilde{\eta}})_{43} \langle  \tilde{\eta}_3^*\rangle + (Y_{\tilde{\chi}})_{43}\langle \tilde{\chi}_3^*\rangle$    &  $0$  &  $M_{31}$ &  $M_{32}$ &   $M_{33}$
\\ \hline
  \end{tabular}
   \label{nlep}
 \caption{The neutral lepton mass matrix $\sqrt{2} M^n_{ij}$ written so as to highlight the seesaw structure.}
 \end{minipage}}
 
 \end{tiny}

\end{table}

\newpage
\begin{acknowledgments}

 G.R. and S. S. thank Natascia Vignaroli for interesting and useful discussions. 
 A.A. is supported by the Talent Scientific Research Program of College of Physics, Sichuan University, Grant No.1082204112427
\& the Fostering Program in Disciplines Possessing Novel Features for Natural Science of Sichuan University,  Grant No. 2020SCUNL209
\& 1000 Talent program of Sichuan province 2021.  
 Work partially supported by Spanish grant PID2020-113775GB-I00 (AEI/ 10.13039/501100011033), Prometeo CIPROM/2021/054 (Generalitat Valenciana),
 by the Government of India, SERB Startup Grant SRG/2020/002303, by MIUR under Project No. 2015P5SBHT and by the INFN research initiative ENP. 
  
\end{acknowledgments}


\bibliographystyle{utphys}
\bibliography{bibliography} 
\end{document}